\documentclass[aps,pra,showpacs,superscriptaddress,preprint]{revtex4}
\usepackage{amsmath}
\usepackage{graphicx}

\catcode`ð=\active
 \defð{\u{g}}
 \catcode`Ð=\active
 \defÐ{\u{G}}
 \catcode`Ý=\active
\defÝ{\. I}
 \catcode`ö=\active
\defö{\"{o}}
 \catcode`Ö=\active
 \defÖ{\"O}
 \catcode`ü=\active
 \defü{\"{u}}
 \catcode`Ü=\active
 \defÜ{\"{U}}
 \catcode`Þ=\active
\defÞ{\c{S}}
 \catcode`þ=\active
 \defþ{\c{s}}
 \catcode`ý=\active
 \defý{{\i}}
 \catcode`ç=\active
\defç{\d{c}}
 \catcode`Ç=\active
\defÇ{\d{C}}

\input{tcilatex}

\begin{document}

\title{Bound state energies and wave functions of spherical quantum dots in
presence of a confining potential model }
\author{Sameer M. Ikhdair}
\email[E-mail: ]{sikhdair@neu.edu.tr}
\affiliation{Physics Department, Near East University, Nicosia, Mersin 10, Turkey}
\date{%
%TCIMACRO{\TeXButton{today}{\today}}%
%BeginExpansion
\today%
%EndExpansion
}

\begin{abstract}
We obtain the exact energy spectra and corresponding wave functions of the
radial Schr\"{o}dinger equation (RSE) for any $(n,l)$ state in the presence
of a combination of psudoharmonic, Coulomb and linear confining potential
terms using an exact analytical iteration method. The interaction potential
model under consideration is Cornell-modified plus harmonic (CMpH) type
which is a correction form to the harmonic, Coulomb and linear confining
potential terms. It is used to investigates the energy of electron in
spherical quantum dot and the heavy quarkonia (QQ-onia).

Keywords: Schr\"{o}dinger equation, confining potentials, spherical quantum
dots, Cornell-modified potential, pseudoharmonic oscillator
\end{abstract}

\pacs{03.65.Fd; 03.65.Ge; 68.65.Hb }
\maketitle

\section{Introduction}

The problem of the inverse-power potential, $1/r^{n},$ has been used on the
level of both classical and quantum mechanics. Some series of inverse power
potentials are applicable to the interatomic interaction in molecular
physics [1-3]. The interaction in one-electron atoms, muonic, hadronic and
Rydberg atoms takes into account inverse-power potentials [4]. Indeed, it
has also been used for the magnetic interaction between spin-$1/2$ particles
with one or more deep wells [5]. The analytical exact solutions of \ this
class of inverse-power potentials, $V(r)=Ar^{-4}+Br^{-3}+Cr^{-2}+Dr^{-1},$ $%
A>0,$ were presented by Barut \textit{et al.} [6] and \"{O}z\c{c}elik and 
\c{S}im\c{s}ek [7] by making an available ansatz for the eigenfunctions. The
Laurent series solutions of the Schr\"{o}dinger equation for power and
inverse-power potentials with two coupling constants $V(r)=Ar^{2}+Br^{-4}$
and three coupling constants $V(r)=Ar^{2}+Br^{-4}+Cr^{-6}$ are obtained
[8,9].

The analytic exact iteration method (AEIM) which demands making a trial
ansatz for the wave function [7] is general enough to be applicable to a
large number of power and inverse-power potentials [10]. Recently, this
method is applied to a class of power and inverse-power confining potentials
of three coupling constants and containing harmonic oscillator, linear and
Coulomb confining terms [11]. This kind of Cornell plus Harmonic (CpH)
confining potential of the form $V(r)=ar^{2}+br-cr^{-1}$ is mostly used to
study individual spherical quantum dots in semiconductors [12] and heavy
quarkonia (QQ-nia) [13,14]. So far, such potentials containing quadratic,
linear and Coulomb terms have been studied [15,16].

The present work considers the the following confining interaction potential
consisting of a sum of pseudoharmonic, linear and Coulombic potential terms: 
\begin{equation}
V(r)=V_{\text{har}}(r)+V_{\text{Corn-}\func{mod}}(r)=ar^{2}+br-\frac{c}{r}-%
\frac{d}{r^{2}},\text{ }a>0,
\end{equation}%
where $a,$ $b,$ $c$ and $d$ are arbitrary constant parameters to be
determined later. The above potential includes the well-known funnel or
Cornell potential, i.e., a Coulomb plus Linear static potential (CpH), $V_{%
\text{Corn}}(r)=br-c/r$ [13], and a term $-d/r^{2}$ is incorporated into the
quarkonium potential for the sake of coherence [14]. We will refer to the
potential model (1) as a Cornell-modified plus harmonic (CMpH) potential,
since the functional form has been improved by the additional $-d/r^{2}$
piece; besides the contribution from the additional term also alters the
value of $b$ and $c$ [14,17]. The authors of Refs. [14,18] did not consider
the harmonic or power-law as the results are expected to be similar. The
CMpH potential is plotted in Figure 1 for the values of parameters: $a=1$ $%
eV.fm^{-2},$ $b=0.217$ $eV$.$fm^{-1},$ $c=0.400$ $eV.fm$ and $d=0.010$ $%
eV.fm^{2}.$

We will apply the AEIM used in [7,11] to obtain the exact energy eigenvalues
and wave functions of the radial Schr\"{o}dinger radial equation (RSE) for
the\ CMpH potential for any arbitrary $(n,l)$ state.

The paper is structured as follows: In Sect. 2, we obtain the exact energy
eigenvalues and wave functions of the RSE in three-dimensions (3D) for the
confining CMpH potential model by proposing asuitable form for the wave
function. In Sect. 3, we apply our results to an electron in spherical
quantum dot of InGaAs semiconductor. The relevant conclusions are given in
Sect. 4.

\section{Exact solution of RSE for the confining potential model}

The three-dimensional ($3D$) Schr\"{o}dinger equation takes the form [19] 
\begin{equation}
\left[ -\frac{\hbar ^{2}}{2m}\Delta +V(r)\right] \psi (r,\theta ,\varphi
)=E_{nl}\psi (r,\theta ,\varphi ),
\end{equation}%
with%
\begin{equation*}
\Delta =\frac{\partial ^{2}}{\partial r^{2}}+\frac{2}{r}\frac{\partial }{%
\partial r}-\frac{L^{2}(\theta ,\varphi )}{\hbar ^{2}r^{2}},\text{ }
\end{equation*}%
where $m$ is the isotropic effective mass and $E_{nl}$ is the total energy
of the particle. For any arbitrary state, the complete wave function, $\psi
(r,\theta ,\varphi ),$ can be written as%
\begin{equation}
\psi (r,\theta ,\varphi )=\dsum\limits_{n,l}N_{l}\psi _{nl}(r)Y_{lm}(\theta
,\varphi ),
\end{equation}%
where spherical harmonic $Y_{lm}(\theta ,\varphi )$ is the eigenfunction of $%
L^{2}(\theta ,\varphi )$ satisfying%
\begin{equation}
L^{2}(\theta ,\varphi )Y_{lm}(\theta ,\varphi )=l(l+1)\hbar
^{2}Y_{lm}(\theta ,\varphi ),
\end{equation}%
and the radial wave function $\psi _{nl}(r)$ is the solution of the equation 
\begin{equation}
\left( \frac{d^{2}}{dr^{2}}+\frac{2}{r}\frac{d}{dr}-\frac{l(l+1)}{r^{2}}%
\right) \psi _{nl}(r)+\frac{2m}{\hbar ^{2}}\left[ E_{nl}-V(r)\right] \psi
_{nl}(r)=0,
\end{equation}%
where $r$ stands for the relative radial coordinates. The radial wave
function $\psi _{nl}(r)$ is well-behaved at the boundaries (the finiteness
of the solution requires that $\psi _{nl}(0)=\psi _{nl}(r\rightarrow \infty
)=0).$ Now, the transformation%
\begin{equation}
\psi _{nl}(r)=\frac{1}{r}\phi _{nl}(r),
\end{equation}%
reduces Eq. (5) to the simple form

\begin{equation}
\phi _{nl}^{\prime \prime }(r)+\left[ \varepsilon _{n,l}-a_{1}r^{2}-b_{1}r+%
\frac{c_{1}}{r}+\frac{d_{1}-l(l+1)}{r^{2}}\right] \phi _{nl}(r)=0,
\end{equation}%
where $\phi _{nl}(r)$ is the reduced radial wave function and 
\begin{equation}
\varepsilon _{nl}=\frac{2m}{\hbar ^{2}}E_{nl},\text{ }a_{1}=\frac{2m}{\hbar
^{2}}a,\text{ }b_{1}=\frac{2m}{\hbar ^{2}}b,\text{ }c_{1}=\frac{2m}{\hbar
^{2}}c,\text{ }d_{1}=\frac{2m}{\hbar ^{2}}d.
\end{equation}%
The analytic exact iteration method (AEIM) requires making the following
ansatze for the wave function [9],%
\begin{equation}
\phi _{nl}(r)=f_{n}(r)\exp \left[ g_{l}(r)\right] ,
\end{equation}%
with 
\begin{subequations}
\begin{equation}
f_{n}(r)=\left\{ 
\begin{array}{cc}
1, & n=0, \\ 
\underset{i=1}{\overset{n}{\Pi }}\left( r-\alpha _{i}^{(n)}\right) , & \text{
}n=1,2,\cdots ,%
\end{array}%
\right.
\end{equation}%
\begin{equation}
g_{l}(r)=-\frac{1}{2}\alpha r^{2}-\beta r+\delta \ln r,\text{ }\alpha >0,%
\text{ }\beta >0.
\end{equation}%
It is clear that $f_{n}(r)$ are equivalent to the Laguerre polynomials [20].
Substituting Eq. (9) into Eq. (5) we obtain 
\end{subequations}
\begin{equation}
\phi _{nl}^{\prime \prime }(r)=\left( g_{l}^{\prime \prime
}(r)+g_{l}^{\prime 2}(r)+\frac{f_{n}^{\prime \prime }(r)+2g_{l}^{\prime
}(r)f_{n}^{\prime }(r)}{f_{n}(r)}\right) \phi _{nl}(r).
\end{equation}%
and comparing Eq. (11) and Eq. (7) yields%
\begin{equation}
a_{1}r^{2}+b_{1}r-\frac{c_{1}}{r}+\frac{l(l+1)-d_{1}}{r^{2}}-\varepsilon
_{nl}=g_{l}^{\prime \prime }(r)+g_{l}^{\prime 2}(r)+\frac{f_{n}^{\prime
\prime }(r)+2g_{l}^{\prime }(r)f_{n}^{\prime }(r)}{f_{n}(r)}.
\end{equation}%
First of all, for $n=0,$ let us take $f_{0}(r)$ and $g_{l}(r)$ given in Eq.
(10b) to solve Eq. (12),%
\begin{equation}
a_{1}r^{2}+b_{1}r-\varepsilon _{0l}-\frac{c_{1}}{r}+\frac{l(l+1)-d_{1}}{r^{2}%
}=\alpha ^{2}r^{2}+2\alpha \beta r-\alpha \left[ 1+2\left( \delta +0\right) %
\right] +\beta ^{2}-\frac{2\beta (\delta +0)}{r}+\frac{\delta \left( \delta
-1\right) }{r^{2}}.
\end{equation}%
By comparing the corresponding powers of $r$ on both sides of Eq. (13) we
find the following corresponding energy and the restrictions on the
potential parameters, 
\begin{subequations}
\begin{equation}
\alpha =\sqrt{a_{1}},
\end{equation}%
\begin{equation}
\beta =\frac{b_{1}}{2\sqrt{a_{1}}},\text{ }a_{1}>0,
\end{equation}%
\begin{equation}
c_{1}=2\beta \left( \delta +0\right) ,
\end{equation}%
\begin{equation}
\delta =\frac{1}{2}\left( 1\pm l^{\prime }\right) ,\text{ where }l^{\prime }=%
\sqrt{\left( 2l+1\right) ^{2}-\frac{8m}{\hbar ^{2}}d}
\end{equation}%
\begin{equation}
\varepsilon _{0l}=\alpha \left[ 1+2\left( \delta +0\right) \right] -\beta
^{2}.
\end{equation}%
Actually, to have well-behaved solutions of the radial wave function at
boundaries, namely the origin and the infinity, we need to take $\delta $
from Eq. (14d) as 
\end{subequations}
\begin{equation}
\delta =\frac{1}{2}\left( 1+l^{\prime }\right) .
\end{equation}%
Therefore, the lowest (ground) state energy from Eq. (14e) together with
Eqs. (14a)-(14c), Eq. (15) and Eq. (8) is given as follows%
\begin{equation}
E_{0l}=\sqrt{\frac{\hbar ^{2}a}{2m}}\left( 2+l^{\prime }\right) -\frac{%
2mc^{2}}{\hbar ^{2}\left( 1+l^{\prime }\right) ^{2}},
\end{equation}%
where the parameter $c$ of potential (1) should satisfy the following
restriction:%
\begin{equation}
c=\frac{b}{2\sqrt{\frac{2ma}{\hbar ^{2}}}}\left( 1+\sqrt{\left( 2l+1\right)
^{2}-\frac{8m}{\hbar ^{2}}d}\right) .
\end{equation}%
Furthermore, the substitution of $\alpha ,$ $\beta $ and $\delta $ from Eqs.
(14a), (14b) and (15), respectively, together with the parameters given in
Eq. (8) into Eqs. (9) and (10), we finally obtain the following ground state
wave function:%
\begin{equation}
\psi _{0l}(r)=N_{0l}r^{\left( -1+l^{\prime }\right) /2}\exp \left( -\frac{1}{%
2}\sqrt{\frac{2ma}{\hbar ^{2}}}r^{2}-\frac{2mc}{\hbar ^{2}\left( 1+l^{\prime
}\right) }r\right) ,
\end{equation}%
with%
\begin{equation*}
N_{0l}=\frac{1}{\sqrt{\Gamma (l^{\prime })D_{-l^{\prime }}\left( \frac{4mc}{%
\hbar ^{2}\left( 1+l^{\prime }\right) }\sqrt{\frac{\hbar }{2\sqrt{2ma}}}%
\right) }}\left( 2\sqrt{\frac{2ma}{\hbar ^{2}}}\right) ^{l^{\prime }/4}\exp
\left( -\frac{1}{2}\sqrt{\frac{2m}{\hbar ^{2}a}}\frac{mc^{2}}{\hbar
^{2}\left( 1+l^{\prime }\right) ^{2}}\right) ,
\end{equation*}%
where $D_{\nu }(z)$ are the parabolic cylinder functions [21]. It should be
noted that the above solutions are well-behaved at the boundaries, i.e., a
regular solution near the origin could be $\phi _{nl}(r\rightarrow
0)\rightarrow r^{\left( 1+l^{\prime }\right) /2}$ and asymptotically at
infinity as $\phi _{nl}(r\rightarrow \infty )\rightarrow \exp \left( -\alpha
r^{2}-\beta r\right) \rightarrow 0.$ When $b=0$ ($c=0$)$,$ the problem turns
to become the commoly known pseudoharmonic oscillator (p.h.o.) interaction ($%
a=m\omega ^{2}/2$)$,$ and consequently $\alpha =m\omega ,$ $\beta =b/\omega $
and $c=(b\delta /m\omega )$ yielding $E_{0l}=\left( 2+l^{\prime }\right) 
\frac{\hbar \omega }{2}-\frac{2mc^{2}}{\hbar ^{2}\left( 1+l^{\prime }\right)
^{2}}$ and wave function $\psi _{0l}(r)=N_{l}r^{\left( -1+l^{\prime }\right)
/2}\exp \left( -\frac{1}{2}\frac{m\omega }{\hbar }r^{2}-\frac{2mc}{\hbar
^{2}\left( 1+l^{\prime }\right) }r\right) ,$ where%
\begin{equation*}
N_{0l}=\frac{1}{\sqrt{\Gamma (l^{\prime })D_{-l^{\prime }}\left( \frac{4mc}{%
\hbar ^{2}\left( 1+l^{\prime }\right) }\sqrt{\frac{\hbar }{2m\omega }}%
\right) }}\left( \frac{2m\omega }{\hbar }\right) ^{l^{\prime }/4}\exp \left(
-\frac{mc^{2}}{4\hbar ^{3}\omega \left( 1+l^{\prime }\right) ^{2}}\right) .
\end{equation*}%
The formula (17) is a relationship between parameters of the potential $a,$ $%
b,$ $c$ and $d.$ Therefore, the solutions (16) and (18) are valid for the
potential parameters satisfying the restriction (17). Moreover, the relation
between the potential parameters (17) depends on the orbital quantum number $%
l$ which means that the potential has to be different for different quantum
numbers. In applying the AEIM, the obtained solution for any potential is
always found to be subjected to certain restrictions on potential parameters
as can be traced in other works (see, for example, [7-9,11]).

Secondly, for the first node ($n=1$), using $f_{1}(r)=(r-\alpha _{1}^{(1)})$
and $g_{l}(r)$ from Eq. (10b) to solve Eq. (12),%
\begin{equation*}
a_{1}r^{2}+b_{1}r-\varepsilon _{1l}-\frac{c_{1}}{r}+\frac{l(l+1)-d_{1}}{r^{2}%
}=\alpha ^{2}r^{2}+2\alpha \beta r
\end{equation*}%
\begin{equation}
-\alpha \left[ 1+2\left( \delta +1\right) \right] +\beta ^{2}-\frac{2\left[
\beta \left( \delta +1\right) +\alpha \alpha _{1}^{(1)}\right] }{r}+\frac{%
\delta \left( \delta -1\right) }{r^{2}}.
\end{equation}%
The relations between the potential parameters and the coefficients $\alpha
, $ $\beta ,$ $\delta $ and $\alpha _{1}^{(1)}$ are%
\begin{equation*}
\alpha =\sqrt{a_{1}},\text{ }\beta =\frac{b_{1}}{2\sqrt{a_{1}}},\text{ }%
\delta =\frac{1}{2}\left( 1+l^{\prime }\right) ,\text{ }\varepsilon
_{1l}=\alpha \left[ 1+2\left( \delta +1\right) \right] -\beta ^{2}.
\end{equation*}%
\begin{equation}
c_{1}-2\beta \left( \delta +1\right) =2\alpha \alpha _{1}^{(1)},\text{ }%
\left( c_{1}-2\beta \delta \right) \alpha _{1}^{(1)}=2\delta ,
\end{equation}%
where $c_{1}$ and $\alpha _{1}^{(1)}$ are found from the constraint
relations, 
\begin{subequations}
\begin{equation}
c=\frac{b}{2\sqrt{\frac{2ma}{\hbar ^{2}}}}\left( 2+l^{\prime }\right) +\sqrt{%
\frac{b^{2}}{\frac{8ma}{\hbar ^{2}}}+\frac{\hbar ^{2}}{m}\sqrt{\frac{\hbar
^{2}a}{2m}}\left( 1+l^{\prime }\right) },
\end{equation}%
\begin{equation}
\alpha \alpha _{1}^{(1)2}+\beta \alpha _{1}^{(1)}-\delta =0\text{ }%
\rightarrow \text{ }\alpha _{1}^{(1)}=-\frac{b}{4a}+\sqrt{\frac{b^{2}}{%
16a^{2}}+\frac{\left( 1+l^{\prime }\right) }{2\sqrt{\frac{2ma}{\hbar ^{2}}}}}%
.
\end{equation}%
The energy eigenvalue is 
\end{subequations}
\begin{equation*}
E_{1l}=\sqrt{\frac{\hbar ^{2}a}{2m}}\left( 4+l^{\prime }\right) -\frac{b^{2}%
}{4a},
\end{equation*}%
\begin{equation}
b=2\sqrt{\frac{2ma}{\hbar ^{2}}}\frac{\left( 2+l^{\prime }\right) c}{\left(
1+l^{\prime }\right) \left( 3+l^{\prime }\right) }\left[ 1+\sqrt{1+\left( 
\frac{\hbar ^{2}}{mc^{2}}\sqrt{\frac{\hbar ^{2}a}{2m}}\left( 1+l^{\prime
}\right) -1\right) \frac{\left( 1+l^{\prime }\right) \left( 3+l^{\prime
}\right) }{\left( 2+l^{\prime }\right) ^{2}}}\right] ,
\end{equation}%
and the wave function is%
\begin{equation}
\psi _{1l}(r)=N_{1l}\left( r-\alpha _{1}^{(1)}\right) r^{\left( -1+l^{\prime
}\right) /2}\exp \left( -\frac{1}{2}\sqrt{\frac{2ma}{\hbar ^{2}}}r^{2}-\sqrt{%
\frac{m}{2\hbar ^{2}a}}br\right) ,
\end{equation}%
with%
\begin{equation*}
N_{1l}=\frac{\left( 2\sqrt{\frac{2ma}{\hbar ^{2}}}\right) ^{l^{\prime
}/4}\exp \left( -\frac{1}{16}\sqrt{\frac{2m}{a^{3}}}\hbar ^{3}b^{2}\right) }{%
\sqrt{\left( 2\sqrt{\frac{2ma}{\hbar ^{2}}}\right) ^{-1}\Gamma (l^{\prime
}+2)S_{1}+\alpha _{1}^{(1)2}\Gamma (l^{\prime })S_{2}-2\left( 2\sqrt{\frac{%
2ma}{\hbar ^{2}}}\right) ^{-1/2}\alpha _{1}^{(1)}\Gamma (l^{\prime }+1)S_{3}}%
},
\end{equation*}%
where%
\begin{equation*}
S_{1}=D_{-(l^{\prime }+2)}\left( \sqrt{\frac{\hbar }{2a}\sqrt{\frac{2m}{a}}}%
b\hbar \right) ,\text{ }S_{2}=D_{-l^{\prime }}\left( \sqrt{\frac{\hbar }{2a}%
\sqrt{\frac{2m}{a}}}b\hbar \right) ,\text{ }S_{3}=D_{-(l^{\prime }+1)}\left( 
\sqrt{\frac{\hbar }{2a}\sqrt{\frac{2m}{a}}}b\hbar \right) ,\text{ }
\end{equation*}%
and $\alpha _{1}^{(1)}$ is given in Eq. (21b). If there is a p.h.o.
interaction, the energy becomes%
\begin{equation}
E_{1l}=\left( 4+l^{\prime }\right) \frac{\hbar \omega }{2}-\frac{b^{2}}{%
2m\omega ^{2}},
\end{equation}%
and the wave function%
\begin{equation}
\psi _{1l}(r)=N_{1l}\left( r-\alpha _{1}^{(1)}\right) r^{\left( -1+l^{\prime
}\right) /2}\exp \left( -\frac{1}{2}m\omega r^{2}-\frac{b}{\omega }r\right) ,
\end{equation}%
with%
\begin{equation*}
N_{1l}=\frac{\left( \frac{2m\omega }{\hbar }\right) ^{l^{\prime }/4}\exp
\left( -\frac{1}{4}\frac{\hbar ^{3}b^{2}}{m\omega ^{3}}\right) }{\sqrt{\frac{%
\hbar }{2m\omega }\Gamma (l^{\prime }+2)S_{1}+\alpha _{1}^{(1)2}\Gamma
(l^{\prime })S_{2}-2\alpha _{1}^{(1)}\sqrt{\frac{\hbar }{2m\omega }}\Gamma
(l^{\prime }+1)S_{3}}},
\end{equation*}%
\begin{equation*}
S_{1}=D_{-(l^{\prime }+2)}\left( \sqrt{\frac{2\hbar }{m\omega }}\frac{b\hbar 
}{\omega }\right) ,\text{ }S_{2}=D_{-l^{\prime }}\left( \sqrt{\frac{2\hbar }{%
m\omega }}\frac{b\hbar }{\omega }\right) ,\text{ }S_{3}=D_{-(l^{\prime
}+1)}\left( \sqrt{\frac{2\hbar }{m\omega }}\frac{b\hbar }{\omega }\right) ,%
\text{ }
\end{equation*}%
where 
\begin{equation*}
b=2\frac{m\omega }{\hbar }\frac{\left( 2+l^{\prime }\right) c}{\left(
1+l^{\prime }\right) \left( 3+l^{\prime }\right) }\left[ 1+\sqrt{1+\left( 
\frac{\hbar ^{3}\omega }{2mc^{2}}\left( 1+l^{\prime }\right) -1\right) \frac{%
\left( 1+l^{\prime }\right) \left( 3+l^{\prime }\right) }{\left( 2+l^{\prime
}\right) ^{2}}}\right] ,
\end{equation*}%
and $\alpha _{1}^{(1)}=\frac{\left( l^{\prime }+1\right) }{2m\omega }.$

Following the analytic iteration procedures for the second node $\left(
n=2\right) $ with $f_{2}(r)=(r-\alpha _{1}^{(2)})(r-\alpha _{2}^{(2)})$ and $%
g_{l}(r)$ as defined in Eq. (10b), we obtain%
\begin{equation*}
a_{1}r^{2}+b_{1}r-\varepsilon _{2,l}-\frac{c_{1}}{r}+\frac{l(l+1)-d_{1}}{%
r^{2}}=\alpha ^{2}r^{2}+2\alpha \beta r
\end{equation*}%
\begin{equation}
-\alpha \left[ 1+2\left( \delta +2\right) \right] +\beta ^{2}-\frac{2\left[
\beta \left( \delta +2\right) +\alpha \dsum\limits_{i=1}^{2}\alpha _{i}^{(2)}%
\right] }{r}+\frac{\delta \left( \delta -1\right) }{r^{2}},
\end{equation}%
The relations between the potential parameters and the coefficients $\alpha
, $ $\beta ,$ $\delta ,$ $\alpha _{1}^{(2)}$ and $\alpha _{2}^{(2)}$ are%
\begin{equation*}
\alpha =\sqrt{a_{1}},\text{ }\beta =\frac{b_{1}}{2\sqrt{a_{1}}},\text{ }%
\delta =\frac{1}{2}\left( 1+l^{\prime }\right) ,\text{ }\varepsilon
_{2,l}=\alpha \left[ 1+2\left( \delta +2\right) \right] -\beta ^{2}.
\end{equation*}%
\begin{equation*}
c_{1}-2\beta \left( \delta +2\right) =2\alpha \dsum\limits_{i=1}^{2}\alpha
_{i}^{(2)},\text{ }\left( c_{1}-2\beta \delta \right)
\dsum\limits_{i<j}^{2}\alpha _{i}^{(2)}\alpha _{j}^{(2)}=2\delta
\dsum\limits_{i=1}^{2}\alpha _{i}^{(2)},
\end{equation*}%
\begin{equation}
\left[ c_{1}-2\beta \left( \delta +1\right) \right] \dsum\limits_{i=1}^{2}%
\alpha _{i}^{(2)}=4\alpha \dsum\limits_{i<j}^{2}\alpha _{i}^{(2)}\alpha
_{j}^{(2)}+2\left( 2\delta +1\right) ,
\end{equation}%
The coefficients $\alpha _{1}^{(2)}$ and $\alpha _{2}^{(2)}$ are found from
the constraint relations, 
\begin{subequations}
\begin{equation}
\alpha \dsum\limits_{i=1}^{2}\alpha _{i}^{(2)2}+\beta
\dsum\limits_{i=1}^{2}\alpha _{i}^{(2)}-\left( 2\delta +1\right) =0,
\end{equation}%
\begin{equation}
\delta \dsum\limits_{i=1}^{2}\alpha _{i}^{(2)2}-\left( \beta
\dsum\limits_{i=1}^{2}\alpha _{i}^{(2)}+1\right)
\dsum\limits_{j<k}^{2}\alpha _{j}^{(2)}\alpha _{k}^{(2)}-2\alpha
\dsum\limits_{j<k}^{2}\alpha _{j}^{(2)2}\alpha _{k}^{(2)2}=0.
\end{equation}%
Hence, the energy eigenvalue is 
\end{subequations}
\begin{equation}
E_{2l}=\sqrt{\frac{\hbar ^{2}a}{2m}}\left( 6+l^{\prime }\right) -\frac{b^{2}%
}{4a},
\end{equation}%
and the associated wave function is%
\begin{equation}
\psi _{2l}(r)=N_{l}\underset{i=1}{\overset{2}{\Pi }}\left( r-\alpha
_{i}^{(2)}\right) r^{\left( -1+l^{\prime }\right) /2}\exp \left( -\frac{1}{2}%
\sqrt{\frac{2ma}{\hbar ^{2}}}r^{2}-\sqrt{\frac{m}{2\hbar ^{2}a}}br\right) ,
\end{equation}%
where $\alpha _{1}^{(2)}$ and $\alpha _{2}^{(2)}$ should satisfy the
restriction relations (28a) and (28b).

We apply the present method for the third node $\left( n=3\right) $ by
taking $f_{3}(r)=(r-\alpha _{1}^{(3)})(r-\alpha _{2}^{(3)})(r-\alpha
_{3}^{(3)})$ and $g_{l}(r)$ as defined in Eq. (10b) to obtain%
\begin{equation*}
a_{1}r^{2}+b_{1}r-\varepsilon _{3,l}-\frac{c_{1}}{r}+\frac{l(l+1)-d_{1}}{%
r^{2}}=\alpha ^{2}r^{2}+2\alpha \beta r
\end{equation*}%
\begin{equation}
-\alpha \left[ 1+2\left( \delta +3\right) \right] +\beta ^{2}-\frac{2\left[
\beta \left( \delta +3\right) +\alpha \dsum\limits_{i=1}^{3}\alpha _{i}^{(3)}%
\right] }{r}+\frac{\delta \left( \delta -1\right) }{r^{2}}.
\end{equation}%
The relations between the potential parameters and the coefficients $\alpha
, $ $\beta ,$ $\delta ,$ $\alpha _{1}^{(3)},$ $\alpha _{2}^{(3)}$ and $%
\alpha _{3}^{(3)}$are%
\begin{equation*}
\alpha =\sqrt{a_{1}},\text{ }\beta =\frac{b_{1}}{2\sqrt{a_{1}}},\text{ }%
\delta =\frac{1}{2}\left( 1+l^{\prime }\right) ,\text{ }\varepsilon
_{3,l}=\alpha \left[ 1+2\left( \delta +3\right) \right] -\beta ^{2}.
\end{equation*}%
\begin{equation*}
c_{1}-2\beta \left( \delta +3\right) =2\alpha \dsum\limits_{i=1}^{3}\alpha
_{i}^{(3)},\text{ }\left( c_{1}-2\beta \delta \right)
\dsum\limits_{i<j<k}^{3}\alpha _{i}^{(3)}\alpha _{j}^{(3)}\alpha
_{k}^{(3)}=2\delta \dsum\limits_{i<j}^{3}\alpha _{i}^{(3)}\alpha _{j}^{(3)},
\end{equation*}%
\begin{equation}
\left[ c_{1}-2\beta \left( \delta +2\right) \right] \dsum\limits_{i=1}^{3}%
\alpha _{i}^{(3)}=4\alpha \dsum\limits_{i<j}^{3}\alpha _{i}^{(3)}\alpha
_{j}^{(3)}+3\left( 2\delta +2\right) .
\end{equation}%
The coefficients $\alpha _{1}^{(3)},$ $\alpha _{2}^{(3)}$ and $\alpha
_{3}^{(3)}$ are found from the constraint relation, 
\begin{equation}
\alpha \dsum\limits_{i=1}^{3}\alpha _{i}^{(3)2}+\beta
\dsum\limits_{i=1}^{3}\alpha _{i}^{(3)}-3\left( \delta +1\right) =0,
\end{equation}%
The energy eigenvalue is 
\begin{equation}
E_{3l}=\sqrt{\frac{\hbar ^{2}a}{2m}}\left( 8+l^{\prime }\right) -\frac{b^{2}%
}{4a},
\end{equation}%
and the wave function is%
\begin{equation}
\psi _{3l}(r)=N_{l}\underset{i=1}{\overset{n=3}{\Pi }}\left( r-\alpha
_{i}^{(n)}\right) r^{\left( -1+l^{\prime }\right) /2}\exp \left( -\frac{1}{2}%
\sqrt{\frac{2ma}{\hbar ^{2}}}r^{2}-\sqrt{\frac{m}{2\hbar ^{2}a}}br\right) .
\end{equation}%
We can repeat this iteration procedures several times to write the exact
energies of the CMpH potential for any $n$ state as%
\begin{equation}
E_{nl}=\sqrt{\frac{\hbar ^{2}a}{2m}}\left( 2+2n+l^{\prime }\right) -\frac{%
b^{2}}{4a},
\end{equation}%
and the wave functions is%
\begin{equation}
\psi _{nl}(r)=N_{l}\underset{i=1}{\overset{n}{\Pi }}\left( r-\alpha
_{i}^{(n)}\right) r^{\left( -1+l^{\prime }\right) /2}\exp \left( -\frac{1}{2}%
\sqrt{\frac{2ma}{\hbar ^{2}}}r^{2}-\sqrt{\frac{m}{2\hbar ^{2}a}}br\right) .
\end{equation}%
The relations between the potential parameters and the coefficients $\alpha
, $ $\beta ,$ $\delta ,$ $\alpha _{1}^{(n)},$ $\alpha _{2}^{(n)},\cdots $, $%
\alpha _{n}^{(n)}$ are%
\begin{equation*}
\alpha =\sqrt{a_{1}},\text{ }\beta =\frac{b_{1}}{2\sqrt{a_{1}}},\text{ }%
\delta =\frac{1}{2}\left( 1+l^{\prime }\right) ,\text{ }\varepsilon
_{2,l}=\alpha \left[ 1+2\left( \delta +n\right) \right] -\beta ^{2},
\end{equation*}%
\begin{equation*}
c_{1}-2\beta \left( \delta +n\right) =0,\text{ (}n=0)
\end{equation*}%
\begin{equation*}
c_{1}-2\beta \left( \delta +n\right) =2\alpha \dsum\limits_{i=1}^{n}\alpha
_{i}^{(n)},\text{ }n=1,2,3,\cdots
\end{equation*}%
\begin{equation*}
\left[ c_{1}-2\beta \left( \delta +n-1\right) \right] \dsum\limits_{i=1}^{n}%
\alpha _{1}^{(n)}=n\left( 2\delta +n-1\right) ,\text{ (}n=1)
\end{equation*}%
\begin{equation*}
\text{ }\left[ c_{1}-2\beta \left( \delta +n-1\right) \right]
\dsum\limits_{i=1}^{n}\alpha _{1}^{(n)}=4\alpha \dsum\limits_{i<j}^{n}\alpha
_{i}^{(n)}\alpha _{j}^{(n)}+n\left( 2\delta +n-1\right) ,\text{ }%
n=2,3,4,\cdots
\end{equation*}%
\begin{equation*}
\left[ c_{1}-2\beta \left( \delta +n-2\right) \right] \dsum\limits_{i<j}^{n}%
\alpha _{i}^{(n)}\alpha _{j}^{(n)}=\left( n-1\right) \left( 2\delta
+n-2\right) \dsum\limits_{i=1}^{2}\alpha _{i}^{(2)},\text{ }\left( n=2\right)
\end{equation*}%
\begin{equation*}
\left( \text{ }c_{1}-2\beta \delta \right) \dsum\limits_{i<j<k}^{n}\alpha
_{i}^{(n)}\alpha _{j}^{(n)}\alpha _{k}^{(n)}=2\delta
\dsum\limits_{i<j}^{n}\alpha _{i}^{(n)}\alpha _{j}^{(n)},\text{ }\left(
n=3\right) ,
\end{equation*}%
\begin{equation*}
\left[ \text{ }c_{1}-2\beta \left( \delta +n-2\right) \right]
\dsum\limits_{i<j}^{n}\alpha _{i}^{(n)}\alpha _{j}^{(n)}=\left( n-1\right)
\left( 2\delta +n-2\right) \dsum\limits_{i<j}^{n}\alpha _{i}^{(n)}\alpha
_{j}^{(n)}
\end{equation*}%
\begin{equation}
+4\alpha \dsum\limits_{i<j<k}^{n}\alpha _{i}^{(n)}\alpha _{j}^{(n)}\alpha
_{k}^{(n)},\text{ }n=3,4,5,\cdots ,
\end{equation}%
and so on.

\section{Results and Discussions}

Now, we consider a special case of potential (1) and an application to our
results. For example, when $b=0$ then leads to $c=0$, then we have the p.h.o
potential, i.e., $V_{ph}(r)=\frac{1}{2}m\omega ^{2}r^{2}-\frac{d}{r^{2}},$
hence, the energy difference between the ground state and the excited states
is%
\begin{equation}
\Delta E=E_{1l}-E_{0l}=\left( 4+l^{\prime }\right) \frac{\hbar \omega }{2}%
-\left( 2+l^{\prime }\right) \frac{\hbar \omega }{2}=\hbar \omega ,
\end{equation}%
which can be used to calculate the values of the potential parameters for
the desired system.

We now apply the present results to describe a realistic physical system
called indium gallium arsenide (InGaAs) quantum dot, i.e., a piece of this
material of a spherical form which is considered as a semiconductor composed
of indium, gallium and arsenic [11]. It is used in high-power and
high-frequency $($say, $\omega \sim 10^{15}$ $Hz)$ electronics because of
its superior electron velocity with respect to the more common
semiconductors silicon and gallium arsenide. InGaAs bandgap also makes it
the detector material of choice in optical fiber communication at $1300$ and 
$1550$ $nm$. The gallium indium arsenide (GaInAs) is an alternative name for
InGaAs. In Fig. 2, we plot the ground state electron energy%
\begin{equation}
E_{0l}(\omega )=\left( 2+\sqrt{\left( 2l+1\right) ^{2}-\frac{8m}{\hbar ^{2}}d%
}\right) \frac{\hbar \omega }{2}-\frac{2mc^{2}}{\hbar ^{2}}\left( 1+\sqrt{%
\left( 2l+1\right) ^{2}-\frac{8m}{\hbar ^{2}}d}\right) ^{-2},
\end{equation}%
versus $\omega $ in the interval $2\times 10^{14}\leq \omega \leq 10\times
10^{14}$ $Hz$ taking the value of $c=0.001$ $eV.fm$ and $d=0$ $eV.fm^{2}$
for the cases $l=0$ and $l=1,$ respectively (harmonic, Coulomb and linear
combination terms). In Fig. 3, we take instead the value of the parameter $%
d=0.1$ $eV.fm^{2}$ (pseudoharmonic, Coulomb and linear combination terms)$.$
The effective mass of electron in the InGaAs semiconductor has been chosen
as $m=0.05m_{e}$ and $\hbar =6.5821\times 10^{-16}$ $eV.s.$ It is seen from
Fig. 2 and Fig. 3 how the increase in the value of $\omega $ leads to an
increase in the energy of electron. The flexibility in the adjustment of the
parameter $d$ allows one to fit the spectrum of the desired model properly
(cf. Fig. 2 and Fig. 3). The parameter $d$ should satisfy the condition \ $%
d\leq \left( 2l+1\right) ^{2}\hbar ^{2}/(8m).$ In Fig. 4 we plot the ground
state wave function $\psi _{0,l}(r)$ of the CpH potential for the cases $l=0$
and $l=1,$ respectively, using the values of potential parameter $c=0.001$ $%
eV.nm$ for an electron with effective mass $m=0.05$ $m_{e}$ and frequency $%
\omega =10\times 10^{14}$ $Hz$. Further, in Fig. 5 we plot the ground state
wave function $\psi _{0,l}(r)$ of the CMpH potential for the cases $l=0$ and 
$l=1,$ respectively, using the values of potential parameters $c=0.001$ $%
eV.nm$ and $d=0.01$ $eV.nm^{2}$ for an electron with effective mass $m=0.05$ 
$m_{e}$ and frequency $\omega =10\times 10^{14}$ $Hz$. In Figs. 6 and 7, we
show electron energy as a function of parameter $c$ in the interval $6\times
10^{-2}\leq c\leq 10\times 10^{-2\text{ \ }}eV.nm$ and $d=0.01$ $eV.nm^{2}$
for frequency $\omega =8\times 10^{14}$ $Hz$ and effective mass $m=0.05$ $%
m_{e}$ for the cases $l=0$ and $l=1,$ respectively. From Fig. 5, the
increase in $c$ leads in the decrease in the electron energy in the InGaAs
semiconductor. In Fig. 8, we plot the first excited state electron energy%
\begin{equation*}
E_{1l}(\omega )=\left( 4+l^{\prime }\right) \frac{\hbar \omega }{2}-\frac{%
2\left( 2+l^{\prime }\right) ^{2}}{\left( 1+l^{\prime }\right) ^{2}\left(
3+l^{\prime }\right) ^{2}}\frac{mc^{2}}{\hbar ^{2}}
\end{equation*}%
\begin{equation}
\times \left[ 1+\sqrt{1+\frac{\left( 1+l^{\prime }\right) \left( 3+l^{\prime
}\right) }{\left( 2+l^{\prime }\right) ^{2}}\left( \frac{\hbar ^{3}\omega }{%
2mc^{2}}\left( 1+l^{\prime }\right) -1\right) }\right] ^{2},
\end{equation}%
versus $\omega $ in the interval $2\times 10^{14}\leq \omega \leq 10\times
10^{14\text{ \ }}Hz$ taking the value of $c=0.001$ $eV.fm$ and $d=0$ $%
eV.fm^{2}$ for the cases $l=0$ and $l=1,$ respectively. In Fig. 9, we take
the value of the parameter $d=0.1$ $eV.fm^{2}.$ We remark that the strongly
attractive singular part $-d/r^{2}$ is physically incorporated into the
quarkonium Cornell potential as the first perturbative term for the sake of
coherence to describe the heavy quarkonia (QQ-nia) (see, for example,
[13,14] and the references therein). It also resemles the centrifugal
barrier term $l(l+1)/r^{2}$ in the Schr\"{o}dinger equation. This attractive
term $-d/r^{2}$ together with the h.o. part $ar^{2}$ constitute the
so-called p.h.o. when $b=0$ $(\beta =0)$ in Eq. (14b) leading to $c=0$ in
Eq. (14c)$.$

In Table 1, we calculate the lowest ($n=0$) energy states ($l=0,1$ and $2)$
from Eq. (36) and from the numerical solution of the radial Schr\"{o}dinger
equation (7) using the values of parameters given by Ref. [34] using the
supersymmetry quantum mechanics (SUSYQM). It is clear that the calculated
energy states in the present work are in good agreement with the results
obtained numerically and SUSYQM [34]. The accracy of our numerical results
is $0.0070\%-0.0095\%.$

\section{Conclusions and Outlook}

In this work, we explored the analytical exact solution for the energy
eigenvalues and their associated wave functions of a particle in the field
of Cornell-modified plus harmonic confining potential. We have used the
analytical exact iteration method (AEIM) which required making a trial
ansatz for the wave function. The general equation for the energy
eigenvalues is given by Eq. (36) with some restrictions on the potential
parameters. If one takes $b=0$ then $c=0,$ hence, the potential (1) turns to
the p.h.o. potential with energy eigenvalues: 
\begin{equation}
E_{nl}=\sqrt{\frac{\hbar ^{2}a}{2m}}\left( 2+2n+\sqrt{\left( 2l+1\right)
^{2}-\frac{8m}{\hbar ^{2}}d}\right)
\end{equation}%
and wave functions:%
\begin{equation}
\psi _{nl}(r)=N_{l}\underset{i=1}{\overset{n}{\Pi }}\left( r-\alpha
_{i}^{(n)}\right) r^{\left( -1+l^{\prime }\right) /2}\exp \left( -\frac{1}{2}%
\sqrt{\frac{2ma}{\hbar ^{2}}}r^{2}\right) .
\end{equation}%
The present results in Eqs. (42) and (43) coincide with Eqs. (15) and (16)
of Ref. [22] obtained by the exact polynomial method, Eqs. (72) and (78) of
Ref. [23] obtained by the Nikiforov-Uvarov method and Eqs. (30) and (31) of
Ref. [24] obtained by the wave function ansatz method after setting $%
D_{0}/r_{0}^{2}=a,$ $D_{0}r_{0}^{2}=-d_{0}$ and $2D_{0}=0.$ The model solved
in the present work can be used in modeling the quarkonium [14] perturbed by
the field of p.h.o. or electron confined in spherical quantum dots [11].
Finally, our solution to this confining potential is being considered
important in many different fields of physics, such as atomic and molecular
physics [25,26], particle physics [13,27,28], plasma physics and solid-state
physics [29-33].

\acknowledgments The partial support provided by the Scientific and
Technological Research Council of Turkey is highly appreciated.

\newpage

{\normalsize %center
}

\bigskip

\bigskip {\normalsize %center
}

\baselineskip= 2\baselineskip% double space the text
%\end{document}
\bigskip \newpage

\begin{table}[tbp]
\caption{Lowest $(n=0)$ energy spectra (for $\hbar =m=1$). }%
\begin{tabular}{llllllll}
\tableline\tableline$a$ & $b$ & $c$ & $d$ & $l$ & Numerical & Present & 
SUSYQM [34] \\ 
\tableline$\frac{1}{32}$ & $1$ & $4$ & $0$ & $0$ & -7.618 & $-7.625$ & $%
-7.625$ \\ 
$\frac{1}{32}$ & $1$ & $8$ & $0$ & $1$ & -7.368 & $-7.375$ & $-7.375$ \\ 
$\frac{1}{32}$ & $1$ & $12$ & $0$ & $2$ & -7.120 & $-7.125$ & $-7.125$ \\ 
\tableline &  &  &  &  &  &  & 
\end{tabular}%
\end{table}
\FRAME{ftbpFO}{0.0277in}{0.0277in}{0pt}{\Qct{A plot of the CMpH potential
[see Eq. (1)] with the selected values of parameters: $a=1$ $eV.fm^{-2},$ $%
b=0.217$ $eV$.$fm^{-1},$ $c=0.400$ $eV.fm$ and $d=0.010$ $eV.fm^{2}.$}}{}{%
Figure 1}{}\FRAME{ftbpFO}{0.0277in}{0.0277in}{0pt}{\Qct{The ground state
electron energy in InGaAs semiconductor versus $\protect\omega $ in the
field of CpH potential with $c=0.001$ $eV.nm$ for cases $l=0$ and $l=1,$
respectively$.$}}{}{Figure 2}{}\FRAME{ftbpFO}{0.0277in}{0.0277in}{0pt}{\Qct{%
The ground state electron energy in InGaAs semiconductor versus $\protect%
\omega $ in the field of the CMpH potential with $c=0.001$ $eV.nm$ and $%
d=0.1 $ $eV.nm^{2}$ for the cases $l=0$ and $l=1,$ respectively.}}{}{Figure 3%
}{}

\bigskip

\FRAME{ftbpFO}{0.0277in}{0.0277in}{0pt}{\Qct{Behaviour of the ground state
wave function $\protect\psi _{n=0,l=0}(r)$ (dashed line) and $\protect\psi %
_{n=0,l=1}(r)$ (continuous line) in the field of the CpH potential with the
value of $c=0.001$ $eV.nm$ for an electron with effective mass $m=0.05$ $%
m_{e}$ and frequency $\protect\omega =10\times 10^{14}$ $Hz$ in the InGaAs
semiconductor.}}{}{Figure 4}{}\FRAME{ftbpFO}{0.0277in}{0.0277in}{0pt}{\Qct{%
Behaviour of the ground state wave function $\protect\psi _{n=0,l=0}(r)$
(dashed line) and $\protect\psi _{n=0,l=1}(r)$ (continuous line) of the CMpH
potential with the values of $c=0.001$ $eV.nm$ and $d=0.01$ $eV.nm^{2}$ for
an electron with an effective mass $m=0.05$ $m_{e}$ and frequency $\protect%
\omega =10\times 10^{14}$ $Hz$ in the InGaAs semiconductor.}}{}{Figure 5}{}%
\FRAME{ftbpFO}{0.0277in}{0.0277in}{0pt}{\Qct{Ground state energy of electron
versus $c,$ for the case $l=0,$ $\protect\omega =8\times 10^{14}$ $Hz$ and $%
d=0.01$ $eV.nm^{2}.$}}{}{Figure 6}{}\FRAME{ftbpFO}{0.0277in}{0.0277in}{0pt}{%
\Qct{Ground state energy of electron versus $c,$ for the case $l=1,$ $%
\protect\omega =8\times 10^{14}$ $Hz$ and $d=0.01$ $eV.nm^{2}.$}}{}{Figure 7%
}{}\FRAME{ftbpFO}{0.0277in}{0.0277in}{0pt}{\Qct{The first excited state
electron energy in InGaAs semiconductor versus $\protect\omega $ in the
field of CpH potential with $c=0.001$ $eV.nm$ for cases $l=0$ and $l=1,$
respectively$.$}}{}{Figure 8}{}\FRAME{ftbpFO}{0.0277in}{0.0277in}{0pt}{\Qct{%
The first excited state electron energy in InGaAs semiconductor versus $%
\protect\omega $ in the field of the CMpH potential with $c=0.001$ $eV.nm$
and $d=0.1$ $eV.nm^{2}$ for the cases $l=0$ and $l=1,$ respectively.}}{}{%
Figure 9}{}

\bigskip

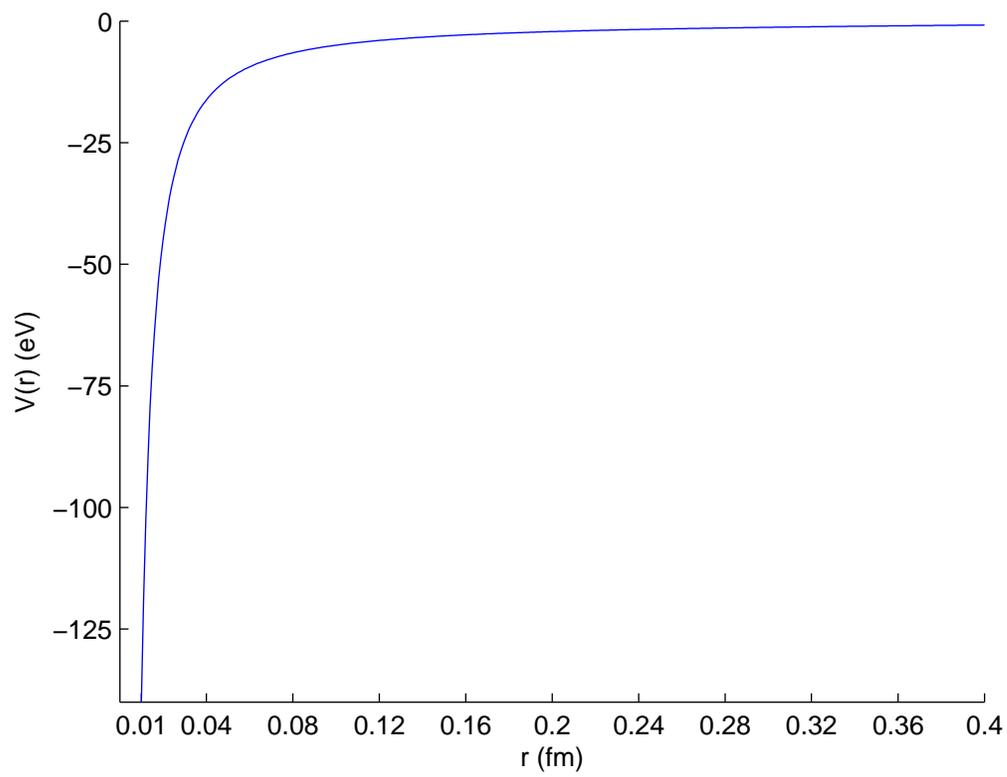
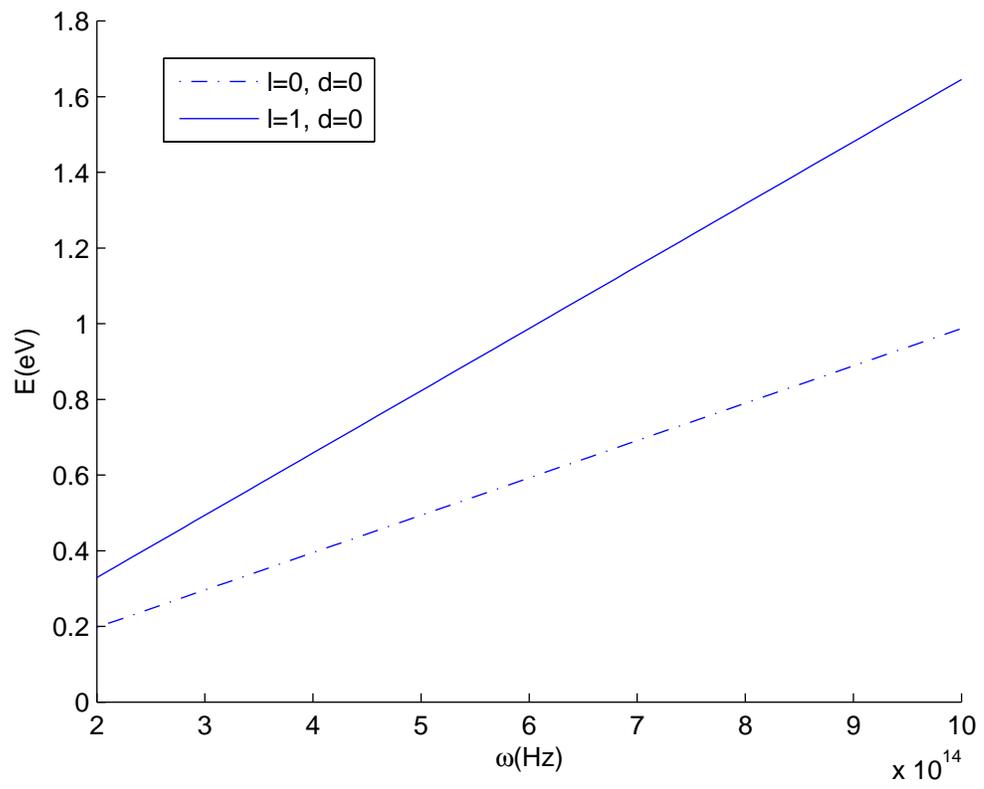
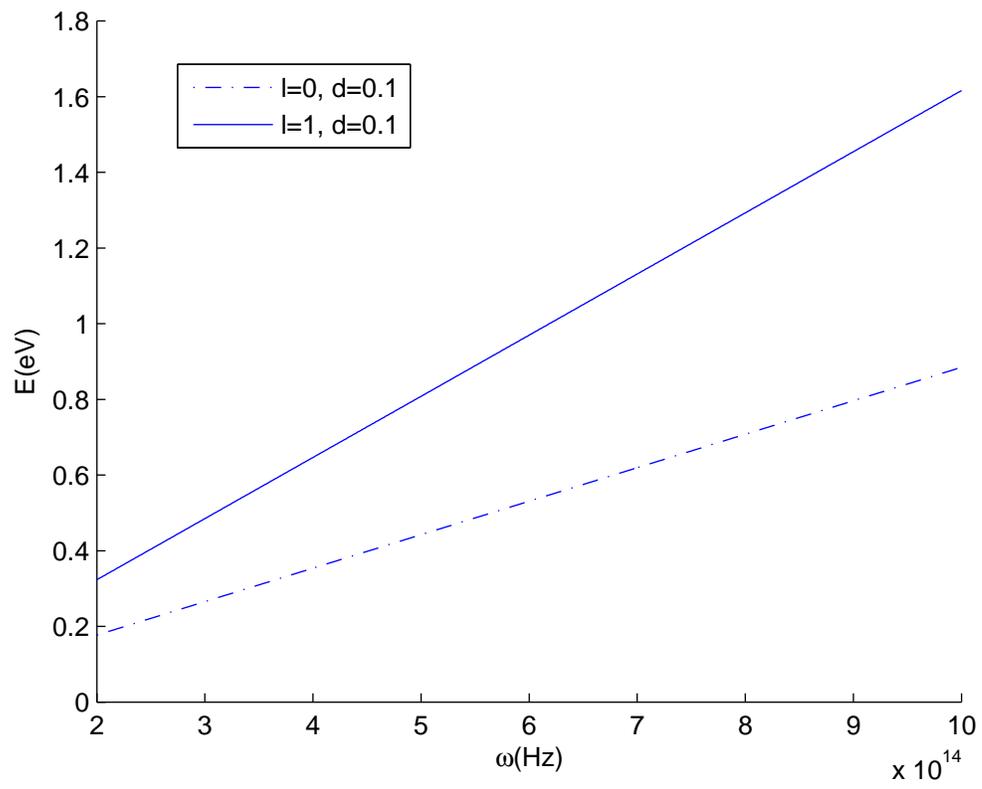
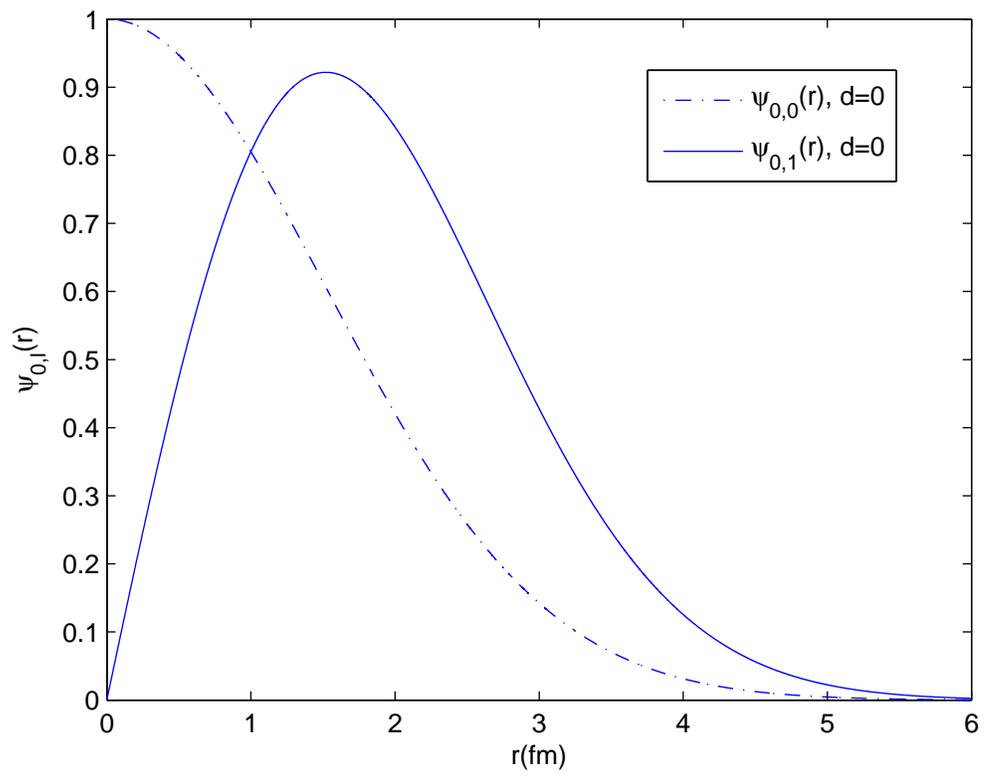
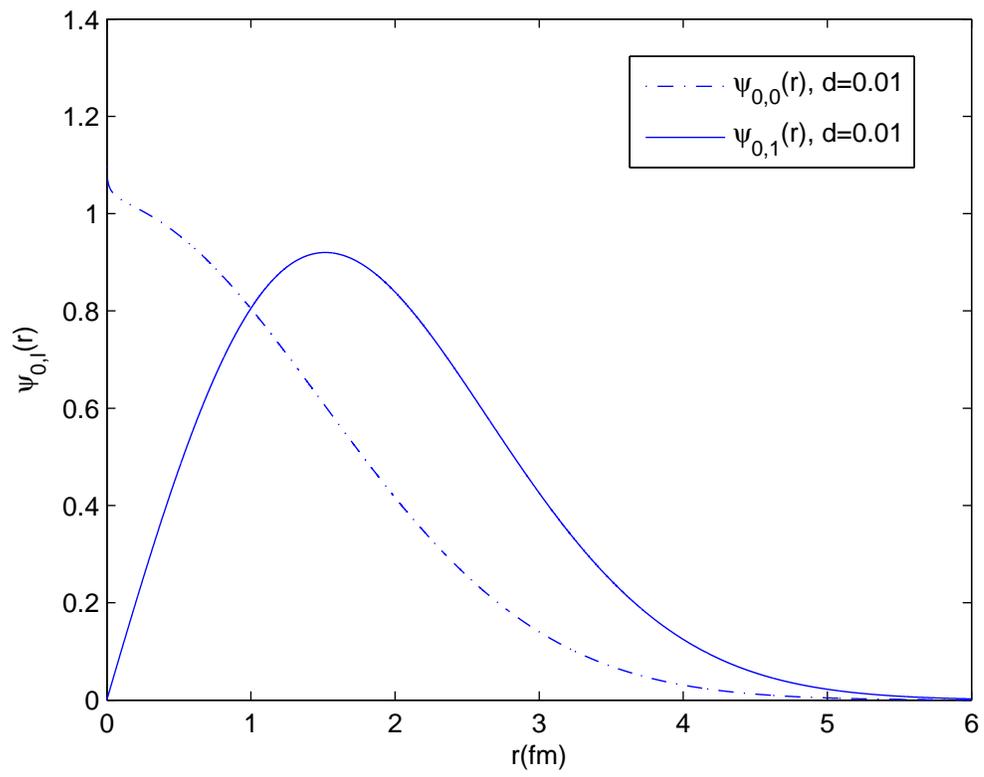
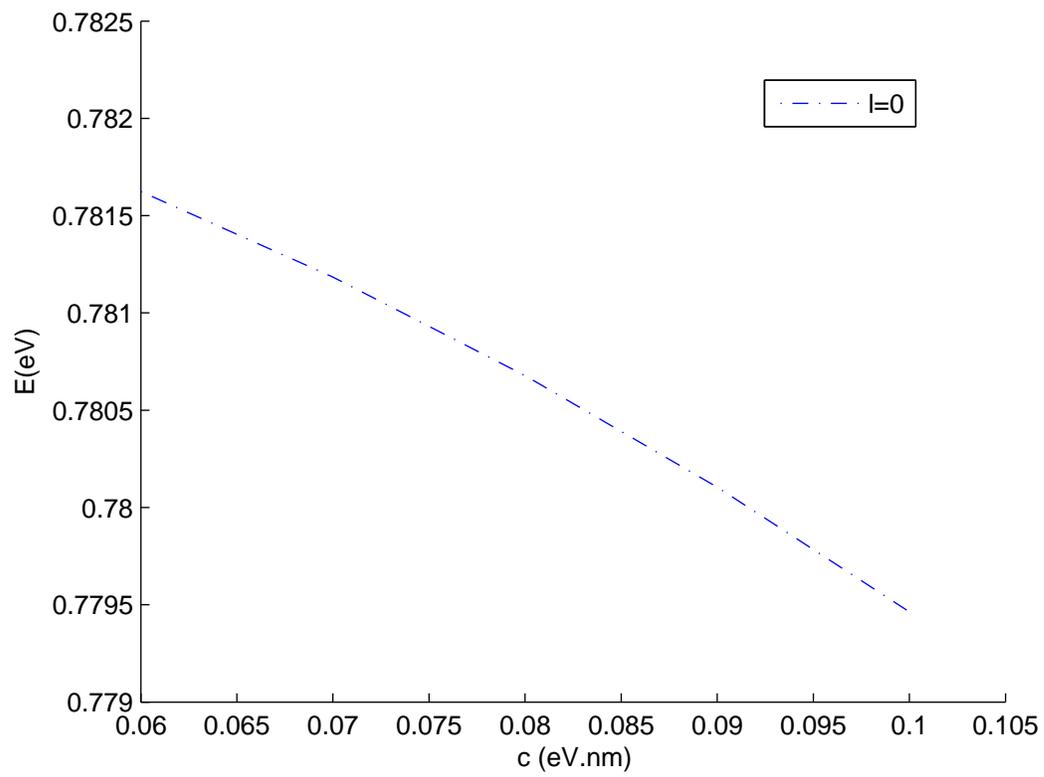
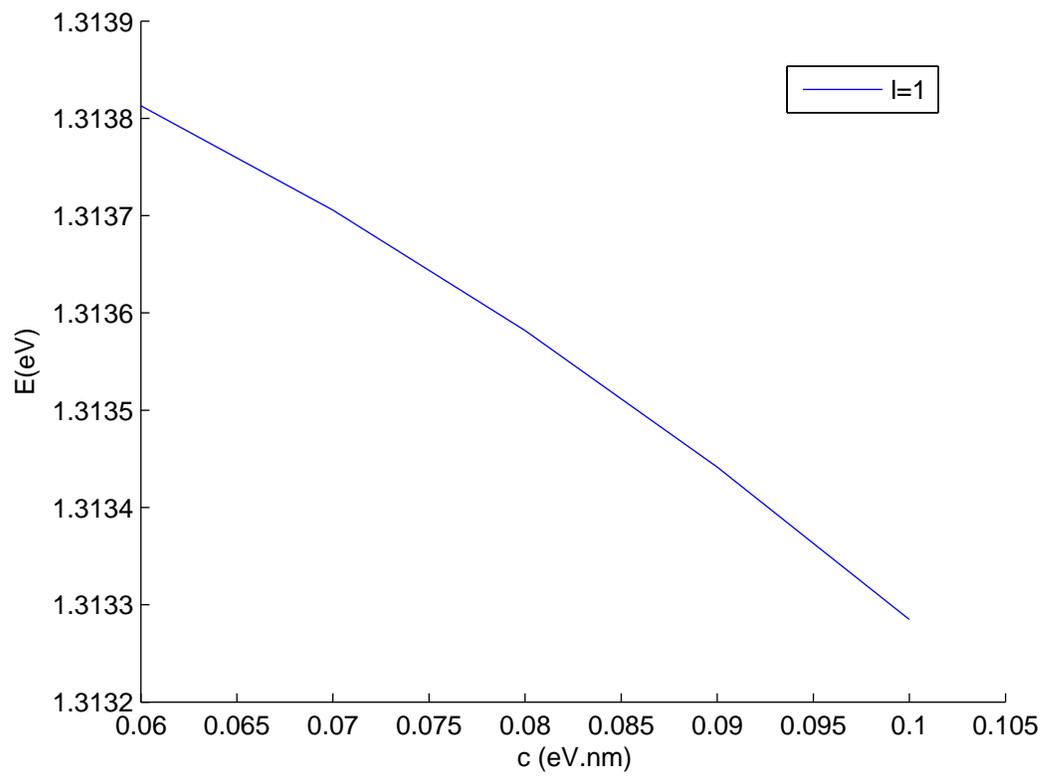
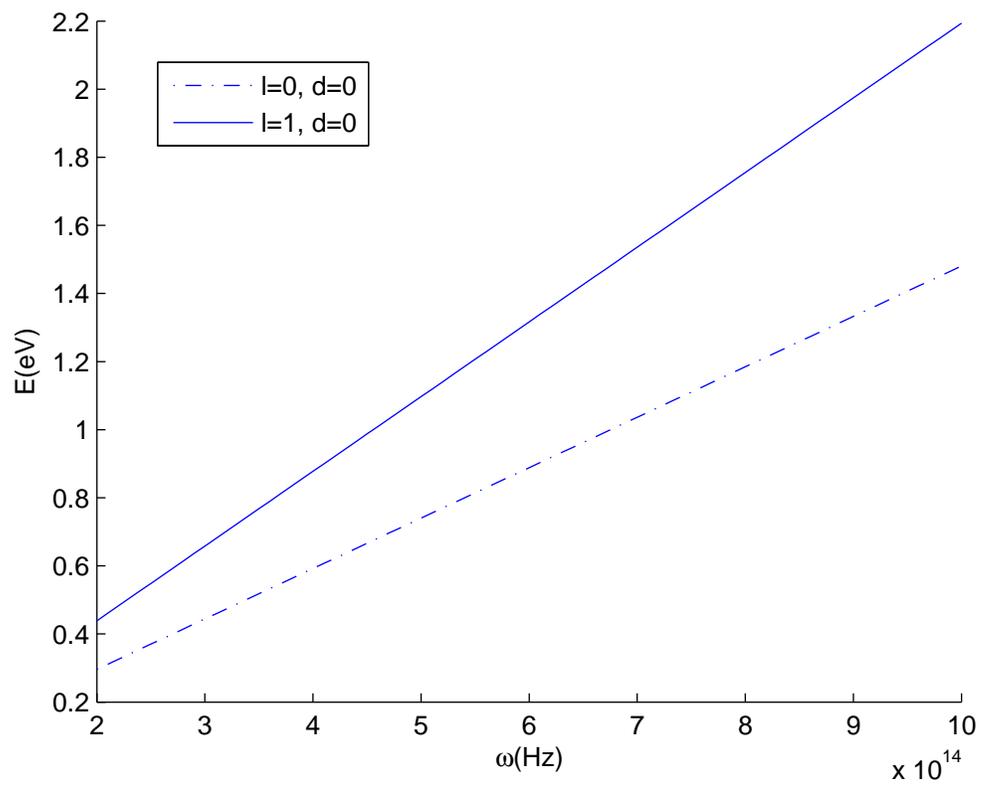
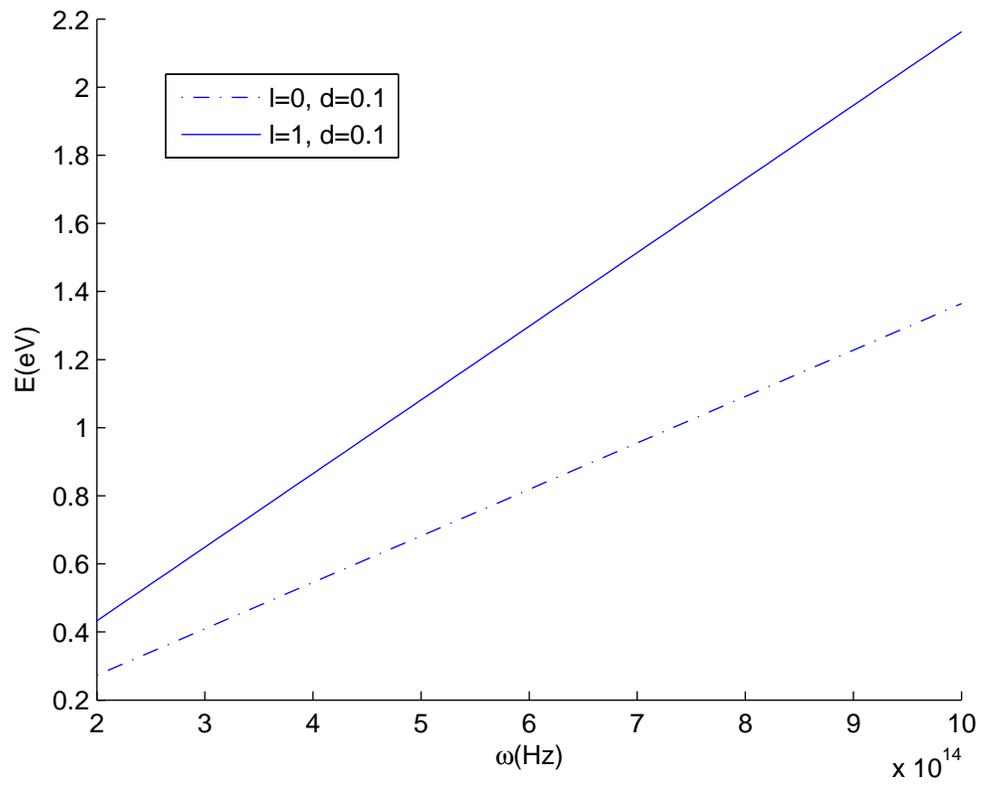
\end{document}